\documentclass[aps,showpacs,preprint,preprintnumber,nofootinbib,amsmath,amssymb,ascmac,12pt]{revtex4}
\usepackage{bm}
\usepackage{color}
\usepackage{graphicx}

\newcommand{\eq}[1]{(\ref{#1})}
\newcommand{\thetah}{\hat{\theta}}
\newcommand{\phih}{\hat{\phi}}
\newcommand{\Mw}{M_{\rm w}}

\newcommand{\Vaf}{V_{\rm 1|after}}

\begin{document}
\title{
Non-linear stability of a brane wormhole
}
\author{
Yumi Akai and Ken-ichi Nakao\footnote{E-mail:knakao@sci.osaka-cu.ac.jp} 
}
\affiliation{
Department of Mathematics and Physics,
Graduate School of Science, Osaka City University,
3-3-138 Sugimoto, Sumiyoshi, Osaka 558-8585, Japan
\vspace{1cm}
}
\begin{abstract}
We analytically study the non-linear stability of a spherically symmetric wormhole 
supported by an infinitesimally thin brane of negative tension, which has been devised by 
Barcelo and Visser. We consider a situation in which a thin spherical shell 
composed of dust falls into an initially static wormhole; The dust shell plays a role of the non-linear disturbance. 
The self-gravity of the falling dust shell is completely taken into account through Israel's formalism 
of the metric junction. When the dust shell goes through the wormhole, 
it necessarily collides with the brane supporting the wormhole. 
We assume the interaction between these shells 
is only gravity and show the condition under which the wormhole stably persists 
after the dust shell goes through it.

\end{abstract}

\preprint{OCU-PHYS-461}
\preprint{AP-GR-136}
\pacs{04.20.-q, 04.20.Gz, 04.70.Bw}

\date{\today}
\maketitle

\section{Introduction}

The wormhole is a fascinating spacetime structure by which shortcut trips 
or travels to disconnected world are possible.  
Active theoretical studies of this subject began by influential papers written by 
Morris, Thorne and Yurtsever\cite{MTY1988} and Morris and Thorne\cite{MT1988}. 
The earlier works are shown in the book written by 
Visser\cite{Visser1995} and review paper by Lobo\cite{Lobo2008}.  

We should note that it is not a trivial task to define a wormhole in mathematically rigorous 
and physically reasonable manner, although we may easily find a wormhole structure in 
each individual case. Hayward gave an elegant definition of the wormhole as an extension of 
the \lq\lq black hole" defined by using trapping horizon\cite{Hayward1999,Hayward2009}. Recently, more sophisticated 
definition has been proposed by 
Tomikawa, Izumi and Shiromizu, and showed that the violation of the 
null energy condition is a necessary condition for the existence of the traversable stationary wormhole 
in the framework of general relativity, where the null energy condition means that  
$T_{\mu\nu}k^\mu k^\nu\geq0$ holds for any null vector 
$k^\mu$  \cite{Tomikawa2015}.\footnote{Several researchers have pointed out an intriguing fact that 
stationary wormhole solutions exist even without the violation 
of the null energy condition, if they have non-vanishing NUT charge which causes closed 
timelike curves\cite{ACZ,GDM}. }

But where does the exotic matter violating the null energy condition appear? 
In Refs.\cite{MTY1988} and \cite{MT1988}, the possibilities of quantum effects were discussed. 
Alternatively, such an exotic matter is often discussed in the context of cosmology. 
The phantom energy, whose pressure $p$ is given through the 
equation of state $p=w\rho$ with $w<-1$ and positive energy
density, $\rho>0$, does not satisfy the null energy condition, and a few researchers showed the 
possibility of the wormhole supported phantom-like matter\cite{Sushkov2005,Lobo2005-1,Lobo2005-2}.  
Recently, theoretical studies from observational point of view on a compact object 
made of the exotic matter, possibly wormholes, have also 
reported\cite{Abe2010,TKAA2011,Nakajima-Asada2012,Tsukamoto-Harada2012,KNA2013}, whereas  
the observational constraint has been reported by Takahashi and Asada\cite{Takahashi-Asada2013}.

It is important to study the stability of wormhole model in order to know whether it is 
really traversable or not. The stability against linear perturbations is a necessary condition 
for the traversable wormhole, but it is insufficient. 
The investigation of non-linear dynamical situation is necessary, 
and there are a few studies 
in this direction\cite{Shinkai-Hayward,Hayward-Koyama2004,Koyama-Hayward2004,NUK2013}. 
{\it In this paper, we also study the non-linear stability of a wormhole in the similar way as that in} 
Ref.\cite{NUK2013}. 

In Ref.~\cite{NUK2013}, the wormhole is assumed to be spherically symmetric and 
be supported by an infinitesimally thin spherical shell. The largest merit of a spherical thin shell 
wormhole is the finite number of its dynamical degrees of 
freedom, and hence we can analyze this model analytically even in highly dynamical cases. 
The thin shell wormhole was first devised by Visser\cite{Visser1989}, and then its stability 
against linear perturbations was investigated by Poisson and Visser\cite{Poisson-Visser1995}. 
Recently, the linear stability of the thin shell wormhole in more general situation has been 
investigated by Garcia, Lobo and Visser\cite{GLV2012}. 

We assume that the spherical shell supporting the wormhole is a 
brane whose equation of state is $P=-\sigma$, where $P$ is the tangential pressure and $\sigma$ is 
the energy per unit area. Furthermore, we assume the existence of spherically symmetric 
electric field. This wormhole model has been devised by Barcelo and Visser\cite{Barcelo-Visser}, and 
its higher dimensional extension has been studied by Kokubu and Harada\cite{Kokubu-Harada}. 
The brane wormhole has a positive gravitational mass; This is 
an important difference between the present study and the previous one in Ref.~\cite{NUK2013} 
in which the gravitational mass of the wormhole is negative.  
The sign of the mass will be significant for the stability, since the positive mass 
may cause the gravitational collapse to form a black hole. 
It is worthwhile to notice that the positivity of the 
mass avoids the observational constraint given in Ref.~\cite{Takahashi-Asada2013}. 
Then as in Ref.~\cite{NUK2013}, we consider a situation in which a infinitesimally thin spherical dust 
shell concentric with the wormhole falls into the wormhole, or in other words,    
plays a role of a non-linear disturbance in the wormhole spacetime. 
These spherical shells are treated by Israel's formulation of metric 
junction\cite{Israel}. When the dust shell goes through the wormhole, it necessarily collides 
with the brane supporting the wormhole. The collision between thin shells has already studied by 
several researchers\cite{NIS1999,IN1999,LMW2002}, and we follow them. 
Then, we show the condition that the wormhole persists after the passage of a spherical shell.  

This paper is organized as follows. In Sec.~II, we derive the equations of motion for 
the brane supporting the wormhole and the spherical dust shell falling 
into the wormhole, in accordance with Israel's formalism of metric junction. 
In Sec.~III, we derive a static solution of the wormhole supported by the brane, which is 
the initial condition. In Sec.~IV, we investigate the condition 
that a dust shell freely falls from infinity and reaches the wormhole throat. 
In Sec.~V, we study the motion of the shells and the change in the gravitational mass 
of the wormhole after collision. In Sec. VI, we show the condition 
that the wormhole persists after the dust shell goes through it. Some complicated 
manipulations and discussions on this subject 
are given separately in Appendix A. Sec. VII is devoted to summary and discussion. 

In this paper, we adopt the geometrized unit in which the speed of light and 
Newton's gravitational constant are one. However, if necessary, they will be recovered. 

\section{Equation of motions for spherical shells}

We consider two concentric spherical shells which are infinitesimally thin. As mentioned 
in the previous section, one is the brane supporting the wormhole and the other is composed of the dust 
which will cause a non-linear perturbation for the wormhole. 

\begin{figure}[b]
\begin{center}
\includegraphics[width=0.3\textwidth]{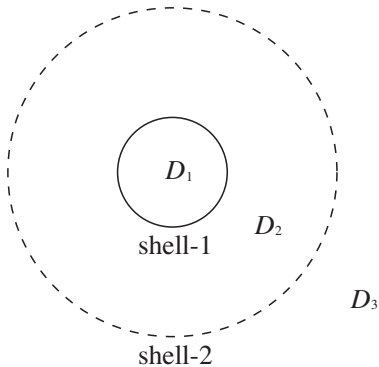}
\caption{\label{initial}
The initial configuration is depicted. 
}
\end{center}
\end{figure}

The trajectories of these shells in the spacetime are timelike hypersurfaces:  
One formed by the brane is denoted by $\Sigma_1$, and the other formed by the dust shell 
is denoted by $\Sigma_2$. These hypersurfaces divide the spacetime into 
three domains denoted by $D_1$, $D_2$ and $D_3$, respectively; $\Sigma_1$ divides the 
spacetime into $D_1$ and $D_2$, whereas $\Sigma_2$ divides the spacetime into 
$D_2$ and $D_3$. 
We also call $\Sigma_1$ and $\Sigma_2$ the shell-1 and the shell-2, respectively. 
This configuration is depicted in Fig.~\ref{initial}. 

The geometry of the domain $D_i$ ($i=1,2,3$) is assumed to be 
described by the Reissner-Nordstr\"{o}m solution: the infinitesimal world interval is given by
\begin{eqnarray}
ds^2=-f_i (r)dt_i^2+\frac{1}{f_i(r)}dr^2
+r^2\left(d\theta^2+\sin^2\theta d\phi^2\right)  \label{RN}
\end{eqnarray}
with
\begin{equation}
f_i(r)=1-\frac{2M_i}{r}+\frac{Q_i^2}{r^2},
\end{equation}
where $M_i$ and $Q_i$ are the mass parameter and the charge parameter, respectively, 
whereas the gauge one-form is given by
\begin{equation}
A_\mu=\left(-\frac{Q_i}{r},0,0,0\right).
\end{equation}
We should note that the coordinate $t_i$ is 
not continuous at the shells, whereas $r$, $\theta$ and $\phi$ are everywhere continuous. 

If $M_i> |Q_i|$ holds, two horizons can exist, and their locations are given by real roots 
of the algebraic equation $f_i(r)=0$:  
\begin{equation}
r=r_{i\pm}:= M_i \pm \sqrt{M_i^2-Q_i^2}.
\end{equation} 
If $M_i=|Q_i|$, there can be one degenerate horizon at $r=M_i$. 
If $M_i<|Q_i|$ holds, the roots of $f_i(r)=0$ are complex or real negative, 
and hence there is no horizon. 

Since finite energy and finite momentum concentrate on the infinitesimally 
thin domains, the stress-energy tensor diverges on these shells. This means 
that these shells are categorized into the so-called curvature polynomial singularity 
through the Einstein equations\cite{Hawking-Ellis}. 
Even though $\Sigma_A$ ($A=1,2$) are spacetime 
singularities, we can derive the equation of 
motion for each spherical shell which is consistent with the Einstein equations 
by so-called Israel's formalism, since each of these singularities is so weak that its 
intrinsic metric exists and the extrinsic curvature defined on each side of $\Sigma_A$ is finite.  

We cover the neighborhood of the singular hypersurface $\Sigma_A$
by a Gaussian normal coordinate $\lambda$, 
where $\partial/\partial\lambda$ is a 
unit vector normal to $\Sigma_A$ and directs from 
$D_A$ to $D_{A+1}$. Then, the sufficient condition 
to apply Israel's formalism is that the stress-energy tensor is written in the form
\begin{equation}
T_{\mu\nu}=S_{\mu\nu}\delta(\lambda-\lambda_A),
\end{equation}
where $\Sigma_A$ is located at $\lambda=\lambda_A$, 
$\delta(x)$ is Dirac's delta function, 
and $S_{\mu\nu}$ is the surface stress-energy tensor on  $\Sigma_A$. 

The junction condition of the metric tensor is obtained as follows.  
We impose that the metric tensor $g_{\mu\nu}$ is continuous even at $\Sigma_A$. 
Hereafter, $n^\mu$ denotes the unit normal vector of $\Sigma_A$, 
instead of $\partial/\partial\lambda$. 
The intrinsic metric of $\Sigma_A$ is given by
\begin{equation}
h_{\mu\nu}=g_{\mu\nu}-n_\mu n_\nu,
\end{equation}
and the extrinsic curvature is defined as
\begin{equation}
K^{(i)}_{\mu\nu}=-h^{\alpha}{}_\mu h^{\beta}{}_\nu \nabla^{(i)}_\alpha n_\beta,
\end{equation}
where $\nabla^{(i)}_\alpha$ is the covariant derivative with respect to the metric in the 
domain $D_i$. This extrinsic curvature describes how $\Sigma_A$ is embedded into 
the domain $D_i$. In accordance with Israel's formalism, the Einstein equations lead to
\begin{equation}
K^{(A+1)}_{\mu\nu}-K^{(A)}_{\mu\nu}=8\pi\left(S_{\mu\nu}
-\frac{1}{2}h_{\mu\nu}{\rm tr}S\right),
\label{j-con-0}
\end{equation}
where ${\rm tr}S$ is the trace of $S_{\mu\nu}$.  Equation (\ref{j-con-0}) gives us the condition 
of the metric junction. 

By the spherical symmetry of the system, the surface stress-energy tensors of the shells 
should be the perfect fluid type; 
\begin{equation}
S_{\mu\nu}=\sigma_A u_\mu u_\nu+P_A(h_{\mu\nu}+u_\mu u_\nu),
\end{equation}
where $\sigma_A$ and $P_A$ are the energy per unit area and the pressure 
on $\Sigma_A$, respectively, and $u^\mu$ is the 4-velocity. 

By the spherical symmetry, 
the motion of the shell-$A$ is described in the form of $t_i=T_{A,i}(\tau)$ and 
$r=R_A(\tau)$, where $i=A$ or $i=A+1$, 
that is to say, $i$ represents one of two domains divided by the shell-$A$, 
and $\tau$ is the proper time of the shell.  
The 4-velocity is given by
\begin{equation}
u^\mu=\left(\dot{T}_{A,i},\dot{R}_A,0,0\right),
\end{equation}
where a dot means the derivative with respect to $\tau$. 
Then, $n_\mu$ is given by
\begin{equation}
n_\mu=\left(-\dot{R}_A,\dot{T}_{A,i},0,0\right).
\end{equation}
Together with $u^\mu$ and $n^\mu$, the following unit vectors form an orthonormal frame;
\begin{eqnarray}
\thetah^\mu&=&\left(0,0,\frac{1}{r},0\right), \\
\phih^\mu&=&\left(0,0,0,\frac{1}{r\sin\theta}\right).
\end{eqnarray}

The extrinsic curvature is obtained as
\begin{eqnarray}
K_{\mu\nu}^{(i)}u^\mu u^\nu&=&\frac{1}{f_{i}\dot{T}_{A,i}}\left(\ddot{R}_A+\frac{f'_i(R_A)}{2}\right), \\
K^{(i)}_{\mu\nu}\thetah^\mu \thetah^\nu&=&
K_{\mu\nu}^{(i)}\phih^\mu \phih^\nu=-n^\mu\partial_\mu \ln r|_{D_i}=-\frac{f_i(R_A)}{R_A}\dot{T}_{A,i}
\label{th-th-comp}
\end{eqnarray}
and the other components vanish, 
where a prime means a derivative with respect to its argument. 
By the normalization condition $u^\mu u_\mu=-1$, we have
\begin{equation}
\dot{T}_{A,i}= \pm\frac{1}{f_i(R_A)}\sqrt{\dot{R}_A^2+f_i(R_A)}~. \label{t-dot}
\end{equation}
Substituting the above equation into Eq.~(\ref{th-th-comp}), we have
\begin{equation}
K_{\mu\nu}^{(i)}\thetah^\mu \thetah^\nu=\mp \frac{1}{R_A}\sqrt{\dot{R}_A^2+f_i(R_A)}.
\end{equation}

From the $u$-$u$ component of Eq.~\eq{j-con-0}, we obtain the following relations.
\begin{equation}
\frac{d(\sigma_A R_A^2)}{d\tau}+P_A\frac{dR_A^2}{d\tau}=0. \label{E-con}
\end{equation}
In the case of the following equation of state
\begin{equation}
P_A=w_A \sigma_A, \label{EOS}
\end{equation}
where $w_A$ is constant, by substituting Eq.~(\ref{EOS}) into Eq.~(\ref{E-con}), we obtain
\begin{equation}
\sigma_A \propto R_A^{-2(w_A+1)}. \label{sigma}
\end{equation}

\subsection{The shell-1: The brane}

As mentioned, we assume that the shell-1 is a brane, i.e., 
$$
w_1=-1.
$$
Without loss of generality, we assume $Q_2\geq0$. Furthermore, we focus on the case of   
$$
Q_2=|Q_1|=Q \geq 0.
$$ 
Since the electric charge of the shell-1 is equal to $Q_2-Q_1$,  the electric charge of the shell-1 is zero 
in the case of $Q_2=Q_1$, whereas the electric charge of the shell-1 may not vanish in the case of 
$Q_2=-Q_1$. As will be shown later, the results in both cases are identical to each other. 

\begin{figure}
\begin{center}
\includegraphics[width=0.5\textwidth]{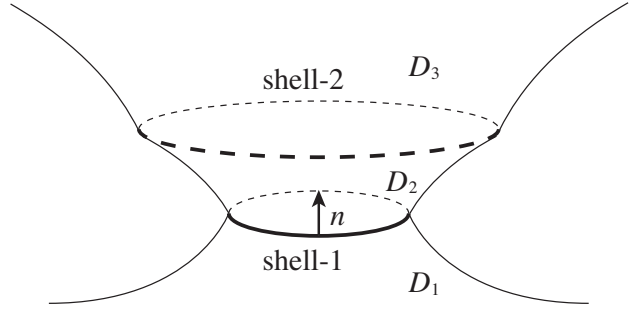}
\caption{\label{fig:wh}
The shell-1 forms the wormhole structure. 
}
\end{center}
\end{figure}

By the assumption, the union of the domains $D_1$ and $D_2$ should have the wormhole 
structure by the shell-1. This means that $n^a\partial_a\ln r |_{D_1}<0$ and 
$n^a\partial_a\ln r |_{D_2}>0$ (see Fig.~\ref{fig:wh}), and we have 
\begin{equation}
K_{\mu\nu}^{(1)}\thetah^\mu \thetah^\nu= +\frac{1}{R_1}\sqrt{\dot{R}_1^2+f_1}
~~~~~{\rm and}~~~~~
K_{\mu\nu}^{(2)}\thetah^\mu \thetah^\nu= -\frac{1}{R_1}\sqrt{\dot{R}_1^2+f_2}.
\label{K-1}
\end{equation}
Here, note that Eq.~(\ref{K-1}) implies $\dot{T}_{1,1}$ is negative, whereas $\dot{T}_{1,2}$ is positive. 
Hence, the direction of the time coordinate basis vector in $D_1$ is opposite with that in $D_2$. 

From $\theta$-$\theta$ component of Eq.~\eq{j-con-0}, we obtain the following relations. 
\begin{equation}
\sqrt{\dot{R}_1^2+f_2(R_1)}+\sqrt{\dot{R}_1^2+f_1(R_1)}= -4\pi \sigma_1 R_1. \label{j-con-1}
\end{equation}
Equation (\ref{j-con-1}) is satisfied only if $\sigma_1$ is negative, 
and hence we assume so. 
 From Eq.~(\ref{sigma}), we have 
\begin{equation}
\sigma_1=-\frac{\mu}{4\pi},
\end{equation}
where $\mu$ is a positive constant, and, hereafter, we call it the stress constant.

Let us rewrite Eq.~\eq{j-con-1} into the form of the energy equation for the shell-1. 
First, we write it in the form
\begin{equation}
\sqrt{\dot{R}_1^2+f_2(R_1)}=-\sqrt{\dot{R}_1^2+f_1(R_1)}+\mu R_1,
\label{j-con-1-1}
\end{equation}
and then take a square of the both sides of the above equation to obtain
\begin{equation}
\sqrt{\dot{R}_1^2+f_1(R_1)}=\frac{1}{2\mu R_1}\left[f_1(R_1)-f_2(R_1)
+\left(\mu R_1\right)^2\right].
\label{j-con-1-3}
\end{equation}
By taking a square of the both sides of the above equation, we have
\begin{equation}
\dot{R}_1^2+V_1(R_1)=0, 
\label{energy-eq}
\end{equation}
where
\begin{align}
V_1(r)
&=1-\frac{1}{r^4}\left(\frac{M_2-M_1}{\mu}\right)^2-\frac{M_1+M_2}{r}+\frac{Q^2}{r^2}-\left(\frac{\mu}{2}\right)^2 r^2.
\label{potential}
\end{align}
Equation (\ref{energy-eq}) is regarded as the energy equation for the shell-1. 
The function $V_1$ corresponds to the effective potential. In the allowed domain 
for the motion of the shell-1, an inequality $V_1\leq0$ should hold. 
But, this inequality is not a sufficient condition of the allowed region. 

The left hand side of Eq.~(\ref{j-con-1-1}) is non-negative, and hence the right hand side of it 
should also be non-negative. Then, substituting Eq.~(\ref{j-con-1-3}) into the right hand side of 
Eq.~(\ref{j-con-1-1}), we have
\begin{eqnarray}
0&\leq& -\sqrt{\dot{R}_1^2+f_1(R_1)}+\mu R_1
=\frac{\mu R_1}{2}-\frac{M_2-M_1}{\mu R_1^2}.
\end{eqnarray}
Further manipulation leads to
\begin{equation}
R_1^3\geq \frac{2}{\mu^2}(M_2-M_1).
\end{equation}
By the similar argument, we obtain
\begin{equation}
-\sqrt{\dot{R}_1^2+f_2(R_1)}+\mu R_1\geq0.
\end{equation}
Then, by the similar procedure, we have
\begin{equation}
R_1^3\geq \frac{2}{\mu^2}(M_1-M_2).
\end{equation}
Hence, we have the following constraint; 
\begin{equation}
R_1\geq \left(\frac{2|M_1-M_2|}{\mu^2}\right)^{{1\over3}}. 
\label{ad-con}
\end{equation}
In order to find the allowed domain for the motion of the shell-1, 
we need to take into account the constraint (\ref{ad-con}) 
in addition to the condition $V_1\leq0$. 

\subsection{The shell-2: The dust shell}

As mentioned, we assume that the shell-2 is composed of non-exotic dust, i.e., $w_2=0$ and $\sigma_2>0$. 
The proper mass of the shell-2 is defined as
\begin{equation}
m_2\equiv 4\pi\sigma_2 R_2^2.
\end{equation}
We find that $m_2$ is constant by Eq.~(\ref{sigma}) and positive by $\sigma_2>0$. 
We also assume 
$$
Q_3=Q_2=Q.
$$ 
This assumption means that the shell-2 is electrically neutral. 

The wormhole structure does not exist around the shell-2 due to $\sigma_2>0$. 
Hence, the extrinsic curvature of the shell-2 is given by
\begin{equation}
K_{\mu\nu}^{(2)}\thetah^\mu \thetah^\nu= -\frac{1}{R_2}\sqrt{\dot{R}_2^2+f_2(R_2)}
~~~~~{\rm and}~~~~~
K_{\mu\nu}^{(3)}\thetah^\mu \thetah^\nu= -\frac{1}{R_2}\sqrt{\dot{R}_2^2+f_3(R_2)}.
\end{equation}
By using the above result, the $\theta$-$\theta$ component of the junction condition leads to
\begin{equation}
\sqrt{\dot{R}_2^2+f_3(R_2)}-\sqrt{\dot{R}_2^2+f_2(R_2)}= -\frac{m_2}{R_2}. \label{j-con-out-1}
\end{equation}

Since $m_2$ is positive, we find from the above equation that $f_2(R_2)> f_3(R_2)$, or equivalently, 
$M_3 > M_2$. From Eq.~(\ref{j-con-out-1}), we have
\begin{equation}
\sqrt{\dot{R}_2^2+f_3(R_2)}=\sqrt{\dot{R}_2^2+f_2(R_2)} -\frac{m_2}{R_2}. \label{j-con-out-2}
\end{equation}
By taking the square of the both sides of Eq.~(\ref{j-con-out-2}), we have
\begin{equation}
\sqrt{\dot{R}_2^2+f_2(R_2)}=\frac{M_3-M_2}{m_2}+\frac{m_2}{2R_2}.\label{j-con-out-4}
\end{equation}

By taking a square of both sides of Eq.~(\ref{j-con-out-4}),  
we obtain an energy equation for the shell-2, 
\begin{equation}
\dot{R}_2^2+V_2(R_2)=0, \label{e-eq-2}
\end{equation} 
where
\begin{equation}
V_2(r)=1-E^2-\frac{2M_{\rm d}}{r}+\frac{Q_1^2}{r^2}-\left(\frac{m_2}{2r}\right)^2, \label{V2-def}
\end{equation}
with
\begin{equation}
E\equiv \frac{M_3-M_2}{m_2}~~~~~~{\rm and}~~~~~~
M_{\rm d}\equiv\frac{1}{2}(M_2+M_3).
\end{equation}
Note that $E$ is a constant which corresponds to the specific energy of the shell-2. 

In the allowed domain for the motion of the shell-2, the effective potential $V_2$ should be non-positive. 
But, as in the case of the shell-1,  it is not a sufficient condition for the allowed domain. 
Since the left hand side of Eq.~(\ref{j-con-out-2}) is non-negative, the following inequality should be 
satisfied. 
\begin{equation}
\sqrt{\dot{R}_2^2+f_2(R_2)} -\frac{m_2}{R_2}\geq0. \label{j-con-out-3}
\end{equation}
Substituting Eq.~(\ref{j-con-out-4}) into the left hand side of Eq.~(\ref{j-con-out-3}), we have 
\begin{equation}
R_2\geq  R_{\rm b}:=\frac{m_2^2}{2(M_3-M_2)}. \label{r-cons-out}
\end{equation}
The above inequality should also be taken into account as a condition for the allowed domain.  

As mentioned, in the case of $M_3\geq Q$, the horizon may appear in the domain $D_3$;  
When the radius $R_2$ of the shell-2 becomes smaller than or equal to 
\begin{equation}
R_{\rm H}:=r_{3+}=M_3+\sqrt{M_3^2-Q^2}, \label{RH-def}
\end{equation}
a black hole including both the wormhole and the shell-2 forms. 
Here it should be noted that Eq.~(\ref{r-cons-out}) is derived by using $u^t_2$ is positive, 
but $u^t_2$ can change its sign within the black hole $R_2<R_{\rm H}$. Hence, if $R_{\rm b}$ 
is smaller than $R_{\rm H}$, Eq.~(\ref{r-cons-out}) looses its validity, and thus the allowed domain 
for the motion of the shell-2 is determined by the only condition $V_2\leq0$. 
The allowed domain for the motion of the shell-2 satisfies $V_2\leq0$ and furthermore 
Eq.~(\ref{r-cons-out}) only if $R_{\rm b}\geq R_{\rm H}$. 

\section{Static wormhole solution}

We consider a situation in which the brane supporting the wormhole is initially static and located at $r=a$. 
Furthermore, we assume that the wormhole is initially mirror symmetric with respect to $r=a$, i.e., 
$f_1(r)=f_2(r)=f(r)$, or equivalently, $M_1=M_2=\Mw$. 
In order that the shell-1 is in a static configuration, its areal radius $R_1=a$ should satisfy 
$V_1(a)=0=V'_1(a)$. Furthermore, $V_1''(a)>0$ should hold so that this structure is stable. 

The condition $V_1(a)=0$ leads to the following relation between the stress constant $\mu$ 
and the areal radius $a$; 
\begin{equation}
\mu^2=\frac{4}{a^2}f(a), \label{V=0}
\end{equation}
whereas, together with the above condition, the condition $V_1'(a)=0$ leads to
\begin{equation}
a^2-3\Mw a+2Q^2=0. \label{dV=0}
\end{equation}
The roots of the above equation are given by
$$
a=a_\pm:=\frac{1}{2}\left(3\Mw\pm\sqrt{9\Mw^2-8Q^2}\right). 
$$
The following inequality should hold so that $a$ is real and positive;
\begin{equation}
\Mw\geq\frac{2\sqrt{2}}{3}Q. \label{L-bound}
\end{equation}
Equation (\ref{L-bound}) implies that $\Mw$ is non-negative. 

Together with Eqs.~(\ref{V=0}) and (\ref{dV=0}), the condition $V''_1(a)>0$ leads to
\begin{equation}
a<\sqrt{2}Q. \label{ddV>0}
\end{equation}
The above condition implies that the charge parameter $Q_i$ cannot vanish so that the areal radius 
$a$ is positive. Since we have
$$
a_\pm-\sqrt{2}Q=\frac{1}{2}\sqrt{3\Mw-2\sqrt{2}Q}
\left(\sqrt{3\Mw-2\sqrt{2}Q}\pm\sqrt{3\Mw+2\sqrt{2}Q}\right),
$$
$a=a_+$ does not satisfy Eq.~(\ref{ddV>0}), but $a=a_-$ does. 

Since $\mu^2$ should be positive, Eq.~(\ref{V=0}) implies that $f(a_{-})>0$ should be satisfied. 
By using Eq.~(\ref{dV=0}), the condition $f(a_-)>0$ leads to
$$
3\Mw^2-2Q^2>\Mw\sqrt{9\Mw^2-8Q^2}.
$$
By taking the square of both sides of the above inequality, we obtain $\Mw<Q$. 

To summarize this section,  the areal radius $a$ and the stress constant $\mu$ 
of the static wormhole are given as a function of $\Mw$ and $Q$;   
\begin{align}
a&=\frac{1}{2}\left(3\Mw-\sqrt{9\Mw^2-8Q^2}\right), \label{a-value}\\ 
\mu&=\frac{a}{2}\sqrt{1-\frac{2\Mw}{a}+\frac{Q^2}{a^2}}, \label{mu-value}
\end{align}
with a constraint
\begin{equation}
\Mw<Q<\frac{3}{2\sqrt{2}}\Mw. \label{Q-restriction}
\end{equation}
Equations (\ref{a-value}) and (\ref{Q-restriction}) lead to
\begin{equation}
\Mw<a<\frac{3}{2}\Mw.   \label{a-domain}
\end{equation}
 
\section{Can the shell-2 reach the wormhole throat?}

We consider the condition that the shell-2 enters the wormhole supported by the shell-1. 
The allowed domain for the motion of the shell-2 is determined by the conditions 
(\ref{r-cons-out}) and $V_2\leq0$.  
The shell-2 is assumed to come from the spatial infinity. By this assumption, 
$E\geq1$ should be satisfied so that $V_2(r)<0$ for sufficiently large $r$. 

\subsection{The case of $Q\leq m_2/2$}

In this case, $V_2(r)$ is negative for $r\geq a$. It should be noted that, in this case, 
$$
M_3=M_2+Em_2>\Mw+2EQ>Q
$$
is satisfied, and hence $R_{\rm H}$ is real and positive.  
As explained in the paragraph including Eq.~(\ref{RH-def}), since we have 
\begin{align}
R_{\rm H}-R_{\rm b}&=M_3+\sqrt{M_3^2-Q^2}-\frac{m_2^2}{2(M_3-M_2)} \cr
&=\Mw+\frac{m_2(2E^2-1)}{2E}+\sqrt{M_3^2-Q^2} \cr
&>0, \nonumber
\end{align}
the allowed domain for the motion of the shell-2 is determined by the only condition $V_2<0$, 
and hence the shell-2 can reach the wormhole throat $r=a$ in this case.

\subsection{The case of $Q>m_2/2$}

We consider the case of $E=1$ and that of $E>1$, separately. 

\subsubsection{The case of $E=1$}

In this case, the positive real root of $V_2(R_{\rm z})=0$ is given by 
$$
R_{\rm z}=\frac{4Q^2-m_2^2}{4(2\Mw+m_2)}.
$$
The allowed domain for the motion of the shell-2 is $R_2\geq R_{\rm z}$. 
We have
\begin{align}
a-R_{\rm z}&=\frac{1}{2}\left(3\Mw-\sqrt{9\Mw^2-8Q^2}\right)-\frac{4Q^2-m_2^2}{4(2\Mw+m_2)} \cr
&=\frac{1}{2\Mw+m_2}
\left[\left(\sqrt{9\Mw^2-8Q^2}-\frac{2\Mw+m_2}{4}\right)^2-\frac{25}{4}\Mw^2+7Q^2+\frac{3}{16}m_2^2+\frac{5}{4}\Mw m_2\right] \cr
&>\frac{1}{2\Mw+m_2}
\left[\left(\sqrt{9\Mw^2-8Q^2}-\frac{2\Mw+m_2}{4}\right)^2+\frac{3}{4}Q^2+\frac{3}{16}m_2^2+\frac{5}{4}\Mw m_2\right] \cr
&>0,\label{E=1}
\end{align}
where we have used $\Mw<Q$ in Eq.~(\ref{Q-restriction}). The above inequality implies that 
the shell-2 can reach the wormhole throat $r=a$. 

\subsubsection{The case of $E>1$}

In this case, the positive real root of $V_2(R_{\rm z})=0$ is given by 
$$
R_{\rm z}=\frac{1}{E^2-1}\left[-\Mw-\frac{m_2E}{2}+
\sqrt{\left(\Mw+\frac{m_2E}{2}\right)^2+(E^2-1)\left(Q^2-\frac{m_2^2}{4}\right)}\right].
$$
The allowed domain for the motion of the shell-2 is $R\geq R_{\rm z}$. 
We can easily see that $R_{\rm z}\rightarrow 0$ and so $a>R_{\rm z}$,  
in the limit of $E\rightarrow\infty$. 
The derivative of $R_{\rm z}$ with respect to $E$ with $\Mw$, $m_2$ and $Q$ fixed is given by
\begin{align}
\frac{\partial R_{\rm z}}{\partial E} =&\frac{X-Y}
{\left(E^2-1\right)^2
\sqrt{\left(\Mw+\displaystyle{\frac{1}{2}}m_2E\right)^2+(E^2-1)\left(Q^2-\displaystyle{\frac{m_2^2}{4}}\right)}}, 
\end{align}
where 
\begin{align}			
X=&\left(\frac{1}{2}m_2E^2+\frac{1}{2}m_2+2\Mw E\right)
\sqrt{\left(\Mw+\frac{1}{2}m_2E\right)^2+(E^2-1)\left(Q^2-\frac{m_2^2}{4}\right)}, \\
Y=&Q^2E^3+\frac{3}{2}\Mw m_2E^2-\left(Q^2-\frac{1}{2}m_2^2-2\Mw^2\right)E+\frac{1}{2}\Mw m_2.
\end{align}
It is not so difficult to see that $Y$ is positive for $E\geq1$, whereas 
$X$ is trivially positive.  Since we have
$$
Y^2-X^2=\left(Q^2-\frac{1}{4}m_2^2\right)\left(Q^2E^2+Mm_2E+\frac{1}{4}m_2^2\right)(E^2-1)^2>0,
$$
we find 
$$
\frac{\partial R_{\rm z}}{\partial E} <0
$$
for $E>1$. As a result, since, as already shown, $a>R_{\rm z}$ holds for both $E=1$ and $E\rightarrow\infty$, 
we have $a>R_{\rm z}$ even for $E>1$. 

In the case of $M_3\geq Q$,  as already shown in the case of $Q<m_2/2$, 
since $R_{\rm b}<R_{\rm H}$ holds, the allowed domain for the motion of the shell-2 is determined  
by the only condition $V_2\leq0$. Hence, the shell-2 can reach the wormhole throat $r=a$. 

In the case of $M_3<Q$, or equivalently, $\Mw<Q-m_2E$, no horizon forms in $D_3$, and hence 
we need to study whether $R_{\rm z}$ is larger than $R_{\rm b}$. In the case of $E=1$, we have
$$
R_{\rm z}-R_{\rm b}=\frac{4Q^2-4m_2 \Mw -3m_2^2}{4(2\Mw+m_2)}
>\frac{(2Q-m_2)^2}{4(2\Mw+m_2)} >0.
$$
In the case of $E>1$, we have 
\begin{align}
R_{\rm z}-R_{\rm b}&=
\frac{1}{2E(E^2-1)}\Biggl[
-2ME+m_2-2m_2E^2 \cr
&+2E\sqrt{\left(\Mw+\frac{m_2E}{2}\right)^2+(E^2-1)\left(Q^2-\frac{m_2^2}{4}\right)}
\Biggr] \cr
&>\frac{1}{2E(E^2-1)}\Biggl[
-2ME+m_2-2m_2E^2 \cr
&+2E\sqrt{\left(\Mw+\frac{m_2E}{2}\right)^2+(E^2-1)\left\{(\Mw+m_2E)^2-\frac{m_2^2}{4}\right\}}
\Biggr] \cr
&=\frac{1}{2E(E^2-1)}\left[2ME(E-1)+m_2(2E^2-1)(E-1)\right] \cr
&>0.
\end{align}
Since now we have $R_{\rm b}<R_{\rm z}$ for $E\geq1$, Eq.~(\ref{r-cons-out}) gives 
no additional constraint on the allowed domain for the motion of the shell-2. 
As a result, the shell-2 can reach the wormhole throat $r=a$ also in $M_3<Q$. 

To summarize this section, the shell-2 reaches the wormhole throat $r=a$ from infinity if it moves 
inward initially. This result is different from the case of the wormhole with the negative mass 
studied in Ref.\cite{NUK2013}: In the negative mass case, $E$ should be larger than 
unity, or in other words, the larger initial ingoing velocity than the present positive mass case 
is necessary so that the shell-2 reaches the wormhole throat, 
since the gravity produced by the wormhole with the negative mass is repulsion.

\section{Collision between the shells}

\begin{figure}
\begin{center}
\includegraphics[width=0.3\textwidth]{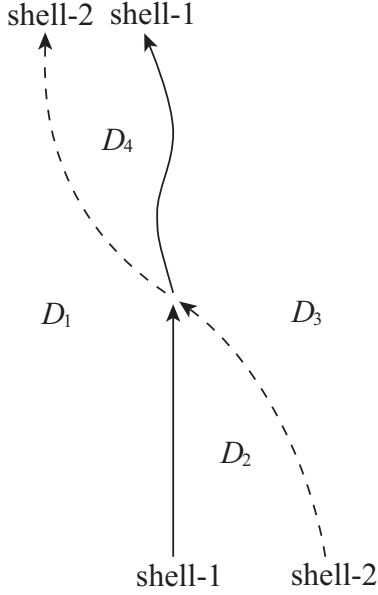}
\caption{\label{collision}
The shell-1 supporting the wormhole is initially static. The shell-2 falls into the wormhole and 
collides with the shell-1. The interaction between these shells is assumed to be gravity only:  
The shells merely go through each other. 
}
\end{center}
\end{figure}

When the shell-2 goes through the wormhole, it necessarily collides with the shell-1 
located at the wormhole throat $r=a$. 
The situation may be recognized by Fig.~\ref{collision}. 
Then, in this section, we show how the 
mass parameter in the domain between the shells changes by the collision.

We assume that the interaction between these shells is gravity only, or in other words, these shells 
merely go through each other: Both of the 4-velocity and the proper mass $4\pi\sigma_AR_A^2$ 
of each shell are continuous at the collision event.

In the domain $D_2$, we may introduce two kinds of the orthonormal frame  
$(u_A^\alpha, n_A^\alpha, \thetah^\alpha, \phih^\alpha)$ at the collision event, where $A=1,2$.  
We can express the 4-velocity $u_1^\alpha$ of the shell-1 by using the orthonormal frame 
$(u_2^\alpha, n_2^\alpha, \theta^\alpha, \phi^\alpha)$, and converse is also possible;
\begin{eqnarray}
u_1^\alpha&=&\left[-u_2^\alpha u_{2\beta}+n_2^\alpha n_{2\beta}+\thetah^\alpha \thetah_\beta
+\phih^\alpha \phih_\beta\right]u_1^\beta
=-(u_1^\beta u_{2\beta})u_2^\alpha+(u_1^\beta n_{2\beta})n_2^\alpha , \\
u_2^\alpha&=&\left[-u_1^\alpha u_{1\beta}+n_1^\alpha n_{1\beta}+\thetah^\alpha \thetah_\beta
+\phih^\alpha \phi_\beta\right]u_2^\beta
=-(u_2^\beta u_{1\beta})u_1^\alpha+(u_2^\beta n_{1\beta})n_1^\alpha.
\end{eqnarray}
The components of $u_A^\alpha$ and $n_A^\alpha$ with respect to the coordinate basis 
in $D_2$ are given by
\begin{eqnarray}
u_1^\alpha&=&\left(\frac{1}{\sqrt{f}},0,0,0\right), \\
n_1^\alpha&=&\left(0,\sqrt{f},0,0\right), \\
u_2^\alpha&=&\left(\frac{1}{f}\sqrt{\dot{R}_2^2+f}, \dot{R}_2,0,0\right), \\
n_2^\alpha&=&\left(\frac{\dot{R}_2}{f},\sqrt{\dot{R}_2^2+f}, 0,0\right), 
\end{eqnarray}
where $f=f(a)$. Hence, we have
\begin{eqnarray}
u_1^\beta u_{2\beta}&=&u_2^\beta u_{1\beta}=-\sqrt{\frac{\dot{R}_2^2}{f}+1}, \\
u_1^\beta n_{2\beta}&=&-\frac{\dot{R}_2}{\sqrt{f}}, \\
u_2^\beta n_{1\beta}&=&\frac{\dot{R}_2}{\sqrt{f}}.
\end{eqnarray}

\subsection{Shell-1 after the collision}

The orthonormal frame $(u_2^\alpha, n_2^\alpha, \thetah^\alpha, \phih^\alpha)$ at the collision event 
is available also in the domain $D_3$. The components of $u_2^\alpha$ and $n_2^\alpha$ 
with respect to the coordinate basis in $D_3$ are given by
\begin{eqnarray}
u_2^\alpha&=&\left(\frac{1}{f_3}\sqrt{\dot{R}_2^2+f_3}, \dot{R}_2,0,0\right), \\
n_2^\alpha&=&\left(\frac{\dot{R}_2}{f_3},\sqrt{\dot{R}_2^2+f_3}, 0,0\right), 
\end{eqnarray}
where $f_3=f_3(a)$. By using the above equations, we obtain 
the components of $u_1^\alpha$ with respect to the coordinate basis in $D_3$ as
\begin{eqnarray}
u_1^{t_3}&=&-(u_1^\beta u_{2\beta})u_2^{t_3}+(u_1^\beta n_{2\beta})n_2^{t_3}
=-(u_1^\beta u_{2\beta})\frac{1}{f_3}\sqrt{\dot{R}_2^2+f_3}
+(u_1^\beta n_{2\beta})\frac{\dot{R}_2}{f_3} \cr
&=&\frac{1}{f_3\sqrt{f}}\left[
\sqrt{(\dot{R}_2^2+f)(\dot{R}_2^2+f_3)}-\dot{R}_2^2
\right], \\
u_1^r&=&-(u_1^\beta u_{2\beta})u_2^{r}+(u_1^\beta n_{2\beta})n_2^{r}
=-(u_1^\beta u_{2\beta})\dot{R}_2+(u_1^\beta n_{2\beta})\sqrt{\dot{R}_2^2+f_3}\cr
&=&\frac{\dot{R}_2}{\sqrt{f}}\left(
\sqrt{\dot{R}_2^2+f}-\sqrt{\dot{R}_2^2+f_3}
\right), \label{ur1}\\
u_1^\theta&=&u_1^\phi=0.
\end{eqnarray}
The above components are regarded as those of the 4-velocity of the shell-1 just after the collision event.  
By using Eqs.~(\ref{j-con-out-1}) and (\ref{ur1}), we have
\begin{equation}
u^r_1=\frac{m_2\dot{R}_2}{a\sqrt{f}}.
\end{equation}
By taking the square of Eq.~(\ref{j-con-out-1}) and using Eq.~(\ref{e-eq-2}), we have
\begin{eqnarray}
\sqrt{(\dot{R}_2^2+f)(\dot{R}_2^2+f_3)}=\dot{R}_2^2+\frac{f+f_3}{2}
-\frac{1}{2}\left(\frac{m_2}{a}\right)^2 =E^2-\left(\frac{m_2}{2a}\right)^2.
\end{eqnarray}
The above equation implies
\begin{equation}
E^2>\left(\frac{m_2}{2a}\right)^2.
\end{equation}
Then, we have
\begin{equation}
u_1^{t_3}=\frac{1}{f_3\sqrt{f}}\left[
1-\frac{2M_{\rm d}}{a}+\frac{Q^2}{a^2}-\frac{1}{2}\left(\frac{m_2}{a}\right)^2
\right].
\end{equation}
We can check that the normalization condition $-f_3 (u_1^{t_3})^2 +f_3^{-1}(u_1^r)^2=-1$ is satisfied. 

The above result implies that just after the collision, the derivative of the areal radius 
of the shell-1 with respect to its proper time becomes
\begin{equation}
\dot{R}_1|_{\rm after}=\frac{m_2\dot{R}_2}{a\sqrt{f}}.
\label{R_1-after}
\end{equation}
Since the shell-2 falls into the wormhole just before the collision, $\dot{R}_2$ is negative. 
This fact implies that the shell-1 or equivalently the radius of the wormhole throat 
begins shrinking just after the collision since $m_2$ is assumed to be positive.  

The domain between the shell-1 and the shell-2 after the collision is called $D_4$. 
From the junction condition between $D_3$ and $D_4$, the shell-2 obeys the following equation 
just after the collision;
\begin{equation}
\dot{R}_1^2|_{\rm after}
=-1+\left(\frac{M_3-M_4}{\mu R_1^2 }\right)^2+\frac{M_3+M_4}{R_1}-\frac{Q^2}{R_1^2}+\left(\frac{\mu R_1}{2}\right)^2.
\label{E-1-after}
\end{equation}
From the above equation and Eq.~(\ref{R_1-after}), we obtain
\begin{equation}
\dot{R}_2^2=-f\left(\frac{a}{m_2}\right)^2
\left[1-\left(\frac{M_3-M_4}{\mu a^2}\right)^2-\frac{M_3+M_4}{a}+\frac{Q^2}{a^2}-\left(\frac{\mu a}{2}\right)^2\right].
\label{m-eq-1}
\end{equation}
Here note that $\dot{R}_2$ is the value of the shell-2 just before the collision. 

\subsection{Shell-2 after the collision}

Since the orthonormal frame $(u_1^\alpha, n_1^\alpha, \thetah^\alpha, \phih^\alpha)$ is 
available also in the domain $D_1$. 
By using Eqs.~(\ref{th-th-comp}), (\ref{t-dot}) and (\ref{K-1}), the components of $u_1^\alpha$ and $n_1^\alpha$ 
with respect to the coordinate basis in $D_1$ are given by
\begin{eqnarray}
u_1^\alpha&=&\left(-\frac{1}{\sqrt{f}},0,0,0\right), \\
n_1^\alpha&=&\left(0, -\sqrt{f}, 0,0\right). 
\end{eqnarray}
As already noted just below Eq.~(\ref{K-1}), the time component of $u_1^\alpha$ 
with respect to the coordinate basis in $D_1$ is negative. 

By using the above equations, we obtain 
the components of $u_2^\alpha$ with respect to the coordinate basis in $D_1$ as
\begin{eqnarray}
u_2^{t_1}&=&-(u_2^\beta u_{1\beta})u_1^{t_1}+(u_2^\beta n_{1\beta})n_1^{t_1}
=(u_2^\beta u_{1\beta})\frac{1}{\sqrt{f}}=-\frac{1}{f}\sqrt{\dot{R}_2^2+f}, \\
u_2^r&=&-(u_2^\beta u_{1\beta})u_1^{r}+(u_2^\beta n_{1\beta})n_1^{r}
=-(u_2^\beta n_{1\beta})\sqrt{f}=-\dot{R}_2,
\label{u_2^r}\\
u_2^\theta&=&u_2^\phi=0.
\end{eqnarray}
Since $\dot{R}_2$ is negative, the shell-2 begins expanding after the collision. 
This is a reasonable result because of the wormhole structure. 

By the spherical symmetry, $D_4$ is also described by the Reissner-Nordstr\"{o}m geometry 
with the mass parameter $M_4$ and the unchanged charge parameter $Q$. 
From the junction condition between $D_1$ and $D_4$, we have
\begin{equation}
\dot{R}_2^2|_{\rm after}=-1+\left(\frac{M_1-M_4}{m_2}\right)^2+\frac{M_1+M_4}{R_2}-\frac{Q^2}{R_2^2}
+\left(\frac{m_2}{2R_2}\right)^2.
\end{equation}
From Eq.~(\ref{u_2^r}), since $\dot{R}_2^2$ is unchanged by the collision, we have 
\begin{equation}
\dot{R}_2^2=-1+\left(\frac{M_1-M_4}{m_2}\right)^2+\frac{M_1+M_4}{a}-\frac{Q^2}{a^2}
+\left(\frac{m_2}{2a}\right)^2. \label{m-eq-2}
\end{equation}
Here again note that $\dot{R}_2$ is the value of the shell-2 just before the collision.

\subsection{The mass parameter $M_4$ in $D_4$}

From Eqs.~(\ref{e-eq-2}) and (\ref{V2-def}), we can write $\dot{R}_2^2$ just before the collision 
in the form
\begin{equation}
\dot{R}_2^2=-1+\left(\frac{M_3-M_2}{m_2}\right)^2+\frac{M_2+M_3}{a}-\frac{Q^2}{a^2}
+\left(\frac{m_2}{2a}\right)^2. \label{V2a}
\end{equation}
Then, Eqs. (\ref{m-eq-1}), (\ref{m-eq-2}) and (\ref{V2a}) determine the unknown parameter $M_4$.

Since $M_1=M_2=\Mw$, Eqs.~(\ref{m-eq-2}) and (\ref{V2a}) lead to
\begin{equation}
\left(\frac{\Mw-M_3}{m_2}\right)^2+\frac{\Mw+M_3}{a}=\left(\frac{\Mw-M_4}{m_2}\right)^2+\frac{\Mw+M_4}{a}.
\end{equation}
By solving the above equation with respect to $M_4$, we obtain two roots, 
$M_4=M_3$ and $M_4=2\Mw-M_3-m_2^2/a$.
By using Eqs.~(\ref{V=0}) and (\ref{m-eq-1}), we find that the latter one, i.e.,  
\begin{equation}
M_4=2\Mw-M_3-\frac{m_2^2}{a} \label{M4}
\end{equation}
is a solution we need, where we have used $M_3=\Mw+m_2E$. 
Hence, after the collision, the wormhole does not have the mirror symmetry with respect to $r=a$.

\section{The condition that the wormhole persists}

In this section, we consider the condition that the wormhole stably 
exists after the passage of the shell-2. 
From Eq.~(\ref{E-1-after}), the effective potential of the shell-1 after the collision is given by
\begin{align}
\Vaf(r)
&=\frac{1}{r^4}\left[-\frac{\mu^2}{4}r^6+r^4-(M_3+M_4)r^3+Q^2r^2-\left(\frac{M_4-M_3}{\mu}\right)^2\right] 
\end{align}
By using Eqs.~(\ref{V=0}), (\ref{dV=0}) and (\ref{M4}), we have
\begin{align}
 \mu^2&=\frac{2(a-\Mw)}{a^3}, \label{mu-def}\\ 
 Q^2&=\frac{a}{2}\left(3\Mw-a\right), \\
 M_3+M_4&=2\Mw-\frac{m_2^2}{a}, \label{M+}\\
 M_3-M_4&=2m_2 E+\frac{m_2^2}{a}. \label{M-}
\end{align}
Equations (\ref{mu-def})--(\ref{M-}) imply that the effective potential $\Vaf$ is characterized by 
four parameters, $\Mw$, $a$, $m_2$ and $E$.   
By regarding $\Mw$ as a parameter to determine the unit of length, 
the motion of the wormhole after the passage of the shell-2 is characterized by three parameters 
$a$, $m_2$ and $E$.  

\subsection{No black-hole formation}

First of all, $a>R_{\rm H}$ must be satisfied in the case of $M_3\geq Q$. 
If not, the wormhole is enclosed by an event horizon 
after the shell-2 enters the domain $r\leq R_{\rm H}$, and hence the wormhole 
cannot stably persist. 

The inequality $M_3 \geq Q$ leads to
\begin{equation}
m_2\geq \frac{1}{E}\left(\sqrt{\frac{a(3\Mw-a)}{2}}-\Mw\right), \label{LB-m2}
\end{equation}
whereas the inequality $a>R_{\rm H}$ leads to
\begin{equation}
m_2<\frac{a-\Mw}{4E}. \label{no-BH}
\end{equation}
If Eq.~(\ref{LB-m2}) holds, Eq.~(\ref{no-BH}) should be satisfied. 
It is not so difficult to see that     
$$
\frac{a-\Mw}{4}>\sqrt{\frac{a(3\Mw-a)}{2}}-\Mw,
$$
and hence both of Eqs.~(\ref{LB-m2}) and (\ref{no-BH}) can hold simultaneously. 
In the case of 
$$
m_2 < \frac{1}{E}\left(\sqrt{\frac{a(3\Mw-a)}{2}}-\Mw\right), 
$$
$M_3<Q$ holds, and hence no horizon appears in $D_3$ even if the shell-2 enters the wormhole. 
As a result, the event horizon does not form by the passage of the shell-2 only if the inequality 
(\ref{no-BH}) holds. 

Hereafter, we focus on the following bounded domain in the parameter space $(a,m_2)$; 
\begin{equation}
{\cal D}=\left\{(a,m_2)\biggl|\Mw<a<\frac{3}{2}\Mw~~{\rm and}~~0<m_2<\frac{a-\Mw}{4E}\right\}.
\label{calD}
\end{equation}

\subsection{Allowed domain for the motion of the shell-1}

The allowed domain for the motion of the shell-1 after the 
collision should be restricted in $r>0$ and bounded  
so that the wormhole stably persists. We introduce a function defined as 
\begin{equation}
W(r):=r^4\Vaf(r)
=-\frac{\mu^2}{4}r^6+r^4-(M_3+M_4)r^3+Q^2r^2-\left(\frac{M_4-M_3}{\mu}\right)^2. 
\label{W-def}
\end{equation}
It is easy to see that the function $W(r)$ has a negative minimum at $r=0$. 
Since $W(r)$ has at most five extrema, $W(r)$ should have two non-negative maxima 
and one negative minimum in $r>0$ and one maximum in $r<0$ 
so that there is a bounded domain of $\Vaf<0$ in $r>0$.  

\begin{figure}
\begin{center}
\includegraphics[width=0.7\textwidth]{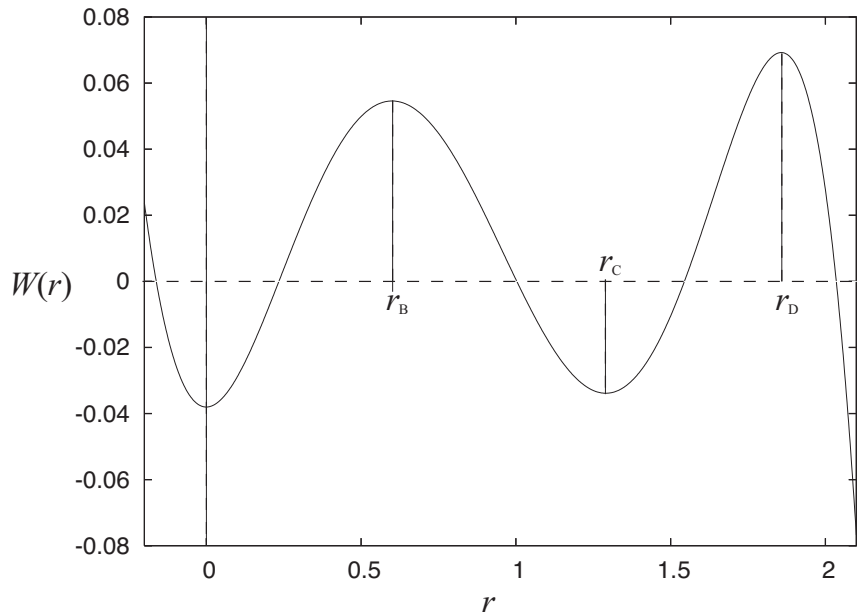}
\caption{\label{after}
Adopting the unit $\Mw=1$, the function $W(r)$ with $(a,m_2,E)=(1.3,0.05,1)$ is depicted. There is a maximum 
of $W(r)$ at $r=r_{\rm A}<0$. However, it is very large compared with extrema in $r\geq0$, and hence we do not show 
it in this figure. }
\end{center}
\end{figure}

We introduce a function $w(r)$ defined as
\begin{equation}
\frac{dW(r)}{dr}=- \frac{3\mu^2}{2} r w(r):=-\frac{3\mu^2}{2} r\left[r^4-\frac{8}{3\mu^2}r^2+\frac{2(M_3+M_4)}{\mu^2}r-\frac{4Q^2}{3\mu^2}\right] .
\end{equation}
The quartic equation $w(r)=0$ should have three positive real roots and one negative real root so that 
there is a bounded domain of $\Vaf<0$ in $r>0$.  
In Appendix A, we see that this is the case as long as the parameters $a$ and $m_2$ are restricted 
to the domain $\cal D$.  
Thus, $W(r)$ has two maxima and one minimum in $r>0$ and one maximum in $r<0$.  
The radial coordinates of the extrema of $W(r)$ other than $r=0$, i.e., the roots of 
$w(r)=0$ are denoted by $r_{\rm A}$, $r_{\rm B}$, $r_{\rm C}$ and $r_{\rm D}$, all of which are the functions of 
not $E$ but $a$ and $m_2$; the explicit forms of $r_{\rm A}$, $r_{\rm B}$, 
$r_{\rm C}$ and $r_{\rm D}$ are given through Ferrari's formula for the roots of a quartic equation, 
but we will not show them here since the expressions of the roots are too complicated to get  
any information from them. We assume $r_{\rm A}<0<r_{\rm B}<r_{\rm C}<r_{\rm D}$, and hence 
$W(r)$ takes maxima at $r=r_{\rm A}$, $r=r_{\rm B}$ and $r=r_{\rm D}$ whereas it takes minima at 
$r=0$ and $r=r_{\rm C}$. (See Fig.~\ref{after}).

\begin{figure}
\begin{center}
\includegraphics[width=0.7\textwidth]{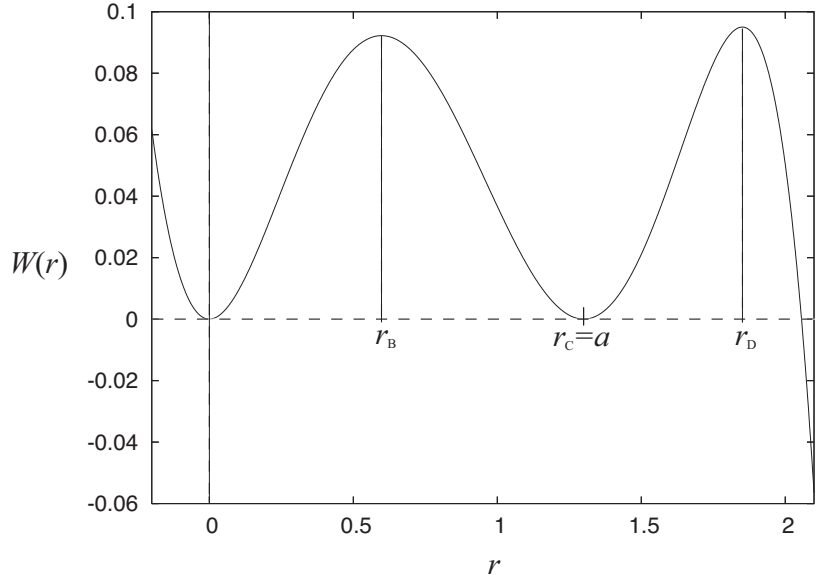}
\caption{\label{static}
The same as Fig.~\ref{after} but $m_2=0$. }
\end{center}
\end{figure}

In the case of $m_2=0$, since $\Vaf(r)$ is equal to $V_1(r)$, we have $r_{\rm C}=a$ and 
$W(r_{\rm C})=0$, and both $W(r_{\rm B})$ and $W(r_{\rm D})$ are positive (see Fig.~\ref{static}). 
By contrast, in the case of non-vanishing $m_2$, we have  
$$
W(a)=-\frac{m_2^2a^2}{2(a-\Mw)}\left[2(2E^2-1)a+2\Mw+4Em_2+\frac{m_2^2}{a}\right]<0,
$$
and hence $W(r_{\rm C})$ must be negative by the continuous dependence of 
$W(r)$ on the parameter $m_2$. 

\begin{figure}
\begin{center}
\includegraphics[width=0.7\textwidth]{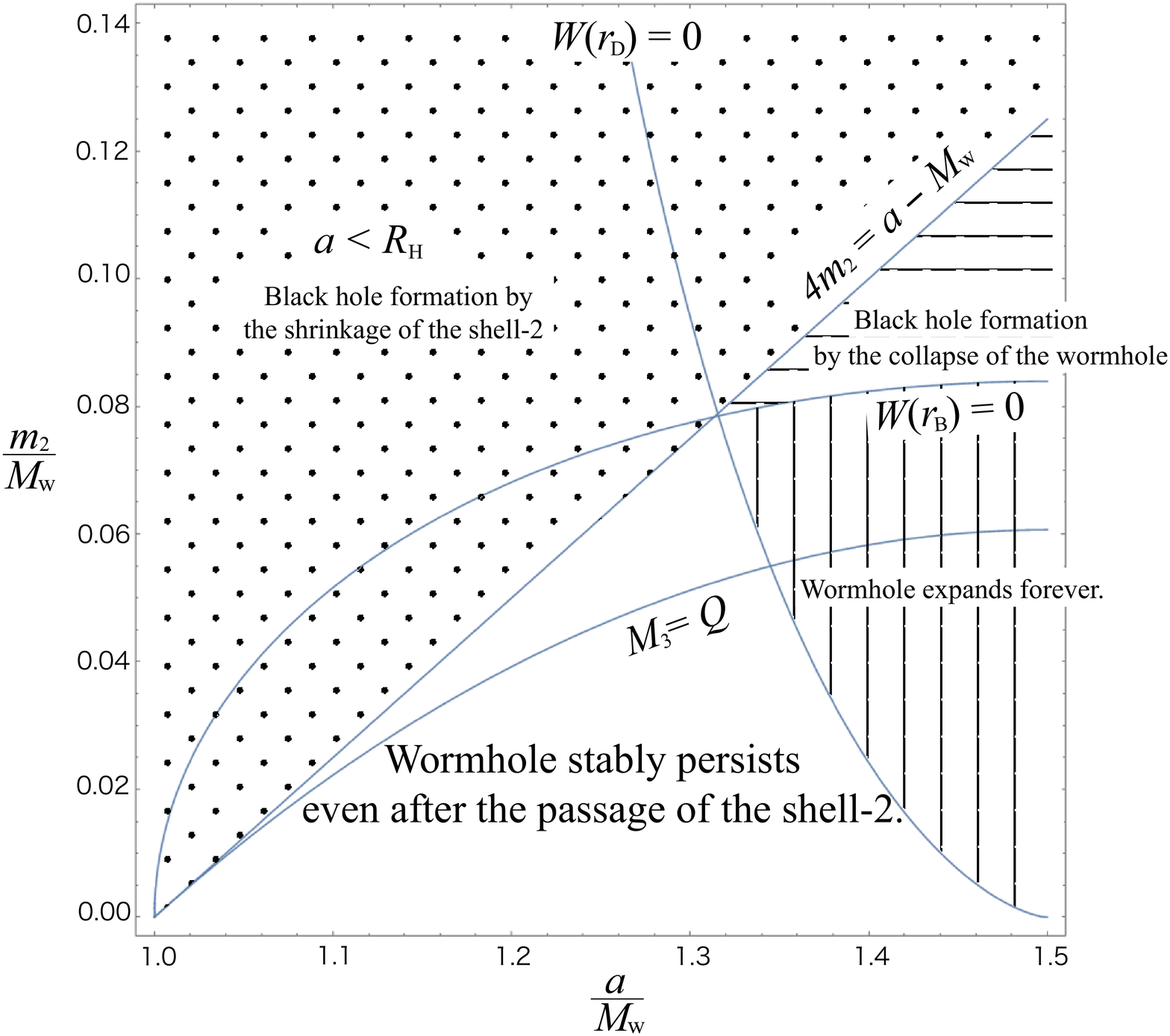}
\caption{\label{fig4}
The $(a,m_2)$-space with $E=1$ is depicted. The domain in which the wormhole stably persists is 
specified as an unshaded region.  
}
\end{center}
\end{figure}

\begin{figure}
\begin{center}
\includegraphics[width=0.7\textwidth]{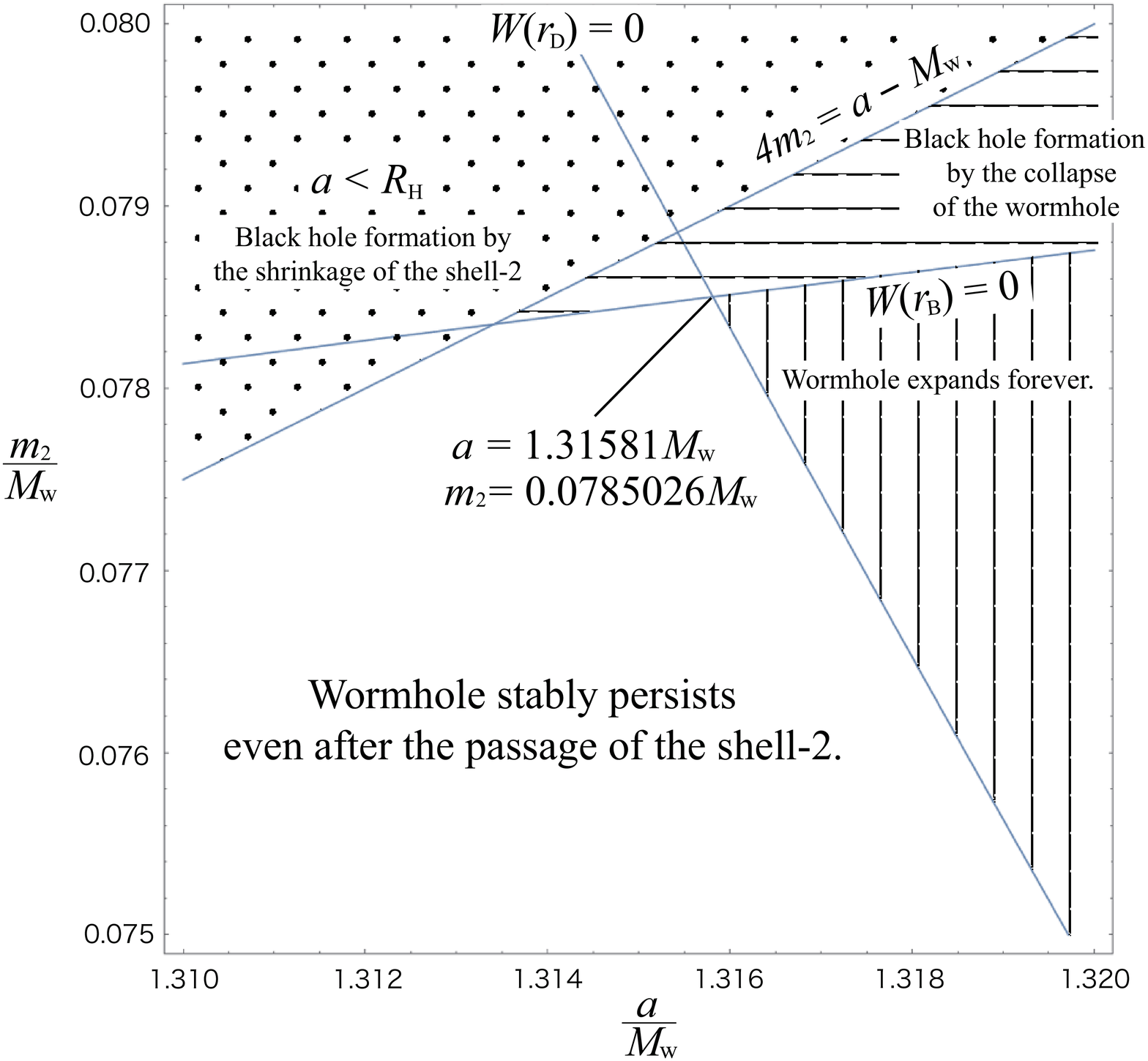}
\caption{\label{fig5}
The close-up of the neighborhood of the intersections of the curves $W(r_{\rm B})=0$, $W(r_{\rm D})=0$ 
and $4m_2=a-\Mw$.  
}
\end{center}
\end{figure}

Since the shell-1 shrinks just after the passage of the shell-2 [see Eq.~(\ref{R_1-after})], 
if $W(r_{\rm B})$ is negative, the shell-1, or equivalently, wormhole collapses to form a black hole. 
If $W(r_{\rm B})$ vanishes, the shell-1 asymptotically approaches $r=r_{\rm B}$ and 
thus the size of the wormhole remains finite. 
If $W(r_{\rm B})$ is positive, the shell-1 bounces off the potential varier, and then 
$R_1$ increases. In this case, $W(r_{\rm D})$ should be equal to or larger than 
zero so that the wormhole persists with its size finite.  
The domain in $(a,m_2)$-space with $E$ fixed in which the wormhole persists after the passage of the 
shell-2 is a curve $W(r_{\rm B})=0$ and a domain restricted by  
$W(r_{\rm B})>0$ and $W(r_{\rm D})\geq0$. 
Hence the critical curves in $(a,m_2)$-space with $E$ fixed are given by the condition
$$
W(r_{\rm B})=0 ~~~{\rm and}~~~W(r_{\rm D})=0.
$$
In Fig.~\ref{fig4}, we depict the domain in $(a,m_2)$-space with $E=1$, in which the wormhole persists after the 
passage of the shell-2, as an unshaded region. 
Figure \ref{fig5} is Fig.~\ref{fig4} in close-up of the neighborhood of the intersections 
of the curves $W(r_{\rm B})=0$, $W(r_{\rm D})=0$ and $a-M=4m_2$, i.e., upper bound of 
the domain $\cal D$. The mass of the shell-2, $m_2$, 
is bounded from above by $0.0785026\Mw$ at which the initial radius of the wormhole throat, $a$, 
equals $1.31581\Mw$. 
This result shows another physically significant difference from the case of the wormhole with the negative mass 
investigated in Ref.~\cite{NUK2013}: The upper bound on $m_2$ is of the order $|\Mw|$ in the negative mass case, 
since the gravitational collapse to form a black hole is prevented by the negative mass of the wormhole. 

\begin{figure}
\begin{center}
\includegraphics[width=0.7\textwidth]{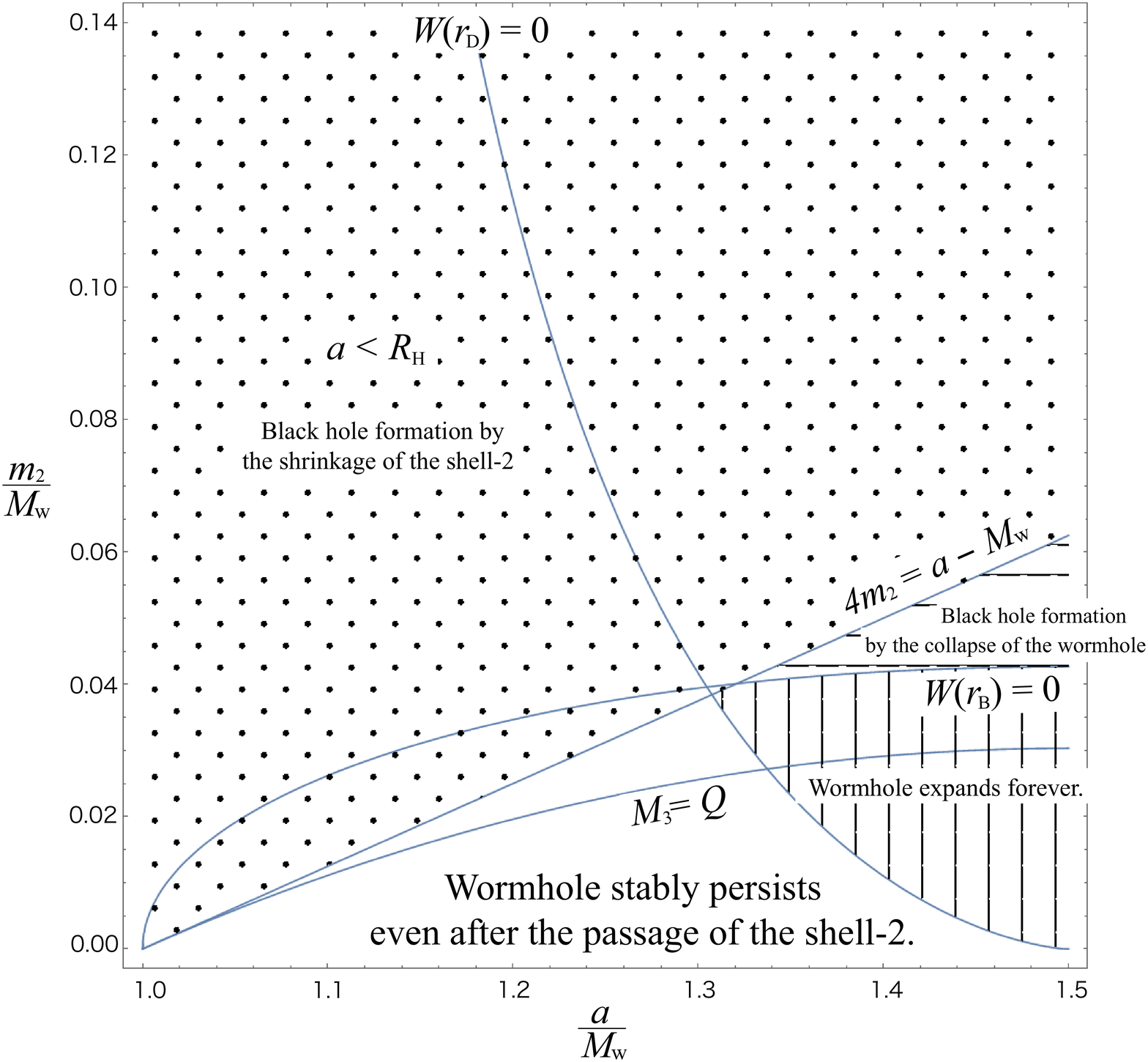}
\caption{\label{fig6}
The same as Fig.\ref{fig4}, but $E=2$. }
\end{center}
\end{figure}

Here it should be noted that $E$ appears only at the last term 
in the right hand side of Eq.~(\ref{W-def}) [see Eq.~(\ref{M-})], and the inclination of $W(r)$ does not depend on $E$.  
The area of the domain in $(a,m_2)$-space in which the wormhole persists 
decreases as $E$ increases. However, there is 
a domain, in which the wormhole persists, for any $E$ larger than unity. 
We depict the same as Fig. \ref{fig4} but $E=2$ in Fig.~\ref{fig6}. 

\section{Summary and discussion}

We analytically studied the non-linear stability of a wormhole supported by an infinitesimally 
thin spherical brane, i.e., a thin spherical shell whose tangential pressure is equal to its energy per unit area with 
an opposite sign; We consider a situation in which a thin spherical shell composed of dust concentric with the brane 
goes through the initially static wormhole in order to play a role of the non-linear disturbance.  
We took into account the self-gravities of both the brane and the dust shell completely through 
Israel's formalism of metric junction. 
The wormhole was assumed to have a mirror symmetry with respect to the brane supporting it.  
As Barcelo and Visser has shown, in such a situation, 
the gravitational mass of the static wormhole should be positive, and 
the static electric field should exist in order that the wormhole is stable 
against linear perturbations. 
Then we studied the condition that the wormhole persists  
after the dust shell goes through it.  
We assumed that the interaction between the brane and the dust shell is only gravity, 
or in other words, the 4-velocities of these shells are assumed to be continuous at the collision event. 
In this model, there are three free parameters;  
The  initial areal radius, $a$, of the wormhole,   
the conserved specific energy $E$ and the proper mass $m_2$, 
of the dust shell, by regarding the initial gravitational mass, $\Mw$, of the wormhole as a unit of length. 
Then, we showed that  
there is a domain of the non-zero measure in $(a, m_2)$-space for $E\geq1$, 
in which the wormhole persists after the dust shell goes through it. In the case of $E=1$, 
the maximum mass of the dust shell $m_2$ is almost equal to $0.08\Mw$. 

Assuming $a\simeq Q\simeq \Mw$, through the geodesic deviation equations, 
the tidal acceleration $A_{\rm tidal}$ felt 
by the spacecraft at the throat of the wormhole, $r=a$, is roughly estimated at
$$
A_{\rm tidal} =\frac{2\Mw\ell}{a^3}\left(\frac{3Q^2}{2\Mw a}-1\right)
\simeq \frac{c^6\ell}{G^2\Mw^2} =10
\left(\frac{\Mw}{4\times 10^5M_\odot}\right)^{-2}
\left(\frac{\ell}{40{\rm m}}\right){\rm m/s}^{2},
$$
where $\ell$ is the length of the spacecraft and $M_\odot$ is the solar mass $2\times10^{30}$kg. 
The area of the wormhole throat with $\Mw=4\times10^5M_\odot$ is about 
$4\pi \Mw^2\simeq 4.5\times10^{12}{\rm km}^2$. Here, let us imagine $10^{12}$ spacecrafts 
placed with almost equal spacing on a sphere concentric with a spherical brane wormhole with $\Mw=4\times10^5M_\odot$. 
The lump of them can be regarded as a dust shell if they are almost freely falling into the wormhole along the radial direction. 
If the size of the spacecraft is about 40m, the tidal acceleration on each spacecraft is the order of 10m/s$^2$ 
even at the throat of the wormhole. Then, since the average separation between 
the nearest spacecrafts is the order of 1km, they can safely go through the wormhole.  
Since the mass of each spacecraft will be about $2\times10^6$kg, 
the total mass of the shell composed of these spacecrafts is $2\times 10^{18}$kg$\simeq 10^{-12}M_\odot$. 
The present result suggests that the wormhole supported by the negative tension brane 
stably persists even after the passage of these spacecrafts.

\section*{Acknowledgments}
We are grateful to Hideki Ishihara, Hirotaka Yoshino and colleagues at the elementary particle physics 
and gravity group in Osaka City University. Especially, YA thanks Hirotaka Yoshino for his advice 
on how to use Mathematica. KN was supported in part by JSPS KAKENHI Grant No.~25400265. 

\appendix
\section{On the roots of the quartic equation $w(r)=0$}

 In this appendix, we show that if the parameters $a$ and $m_2$ are in the domain $\cal D$ 
 of $E\geq1$, the quartic equation $w(r)=0$ has three positive real roots and one negative real root. 
 
 In accordance with Eqs.~(\ref{mu-def})--(\ref{M-}), 
 we regard $\mu$ and $Q$ as functions of $a$, $M_3+M_4$ as 
 a function of $a$ and $m_2$ and $M_3-M_4$ as a function of $a$, $m_2$ and $E$. 

First, we show that $M_3+M_4$ is bounded below by a positive value.    
Because of Eq.~(\ref{no-BH}), by using Eq.~(\ref{M+}), we have 
$$
M_3+M_4=2\Mw-\frac{m_2^2}{a}>N(a,\Mw,E),
$$
where
$$
N(a,\Mw,E):=\frac{(16E^2+1)\Mw}{8E^2}-\frac{1}{16E^2}\left(a+\frac{\Mw^2}{a}\right).
$$
Because of Eq.~(\ref{a-domain}),    
$$
 \frac{\partial N}{\partial a}=\frac{\Mw^2-a^2}{16E^2a^2}<0
$$ 
is satisfied, and hence we have
$$
N(a,\Mw,E)>N\left(\frac{3}{2}\Mw,\Mw,E\right)=\frac{1}{96}\left(192-\frac{1}{E^2}\right)\Mw
\geq\frac{191}{96}\Mw, 
$$
where we have used $E\geq1$. As a result, we obtain
\begin{equation}
M_3+M_4>\frac{191}{96}\Mw. \label{LB-M3M4}
\end{equation}

The derivative of $w(r)$ is given by 
\begin{align}
\frac{dw(r)}{dr}&=4r^3-\frac{16}{3\mu^2}r+\frac{2(M_3+M_4)}{\mu^2}.
\end{align}
If the inequality
\begin{equation}
(M_3+M_4)\mu<\frac{32}{27} \label{UB-muM3M4}
\end{equation}
holds, the cubic equation $dw(r)/dr=0$ has three real roots.  
We show that Eq.~(\ref{UB-muM3M4}) necessarily holds in the domain $\cal D$ of $E\geq1$.  
Because of Eq.~(\ref{a-domain}), we have
\begin{equation}
\frac{d \mu}{d a}=\frac{3\Mw-2a}{a^3}\sqrt{\frac{a}{2(a-\Mw)}}>0, \label{dmu}
\end{equation}
and hence 
\begin{equation}
\mu<\mu|_{a=\frac{3}{2}\Mw}=\frac{1}{\Mw}\sqrt{\frac{8}{27}} \label{UB-mu}
\end{equation}
holds. 
Equation (\ref{UB-mu}) leads to Eq.~(\ref{UB-muM3M4}) as follows;
$$
(M_3+M_4)\mu=\left(2\Mw-\frac{m_2^2}{a}\right)\mu<2\Mw\mu<\sqrt{\frac{32}{27}}<\frac{32}{27}.
$$  

By virtue of Eqs.~(\ref{LB-M3M4}) and (\ref{UB-muM3M4}), we find that 
$w(r)$ has one minimum in $r<0$ and one maximum and one minimum in $r>0$. 

Hereafter, the three real roots  of $dw(r)/dr=0$ are denoted by $r=r_i$ ($i=1,2,3$):  
\begin{equation}
r_1=\frac{4}{3\mu}\cos\left(\frac{\theta}{3}\right), ~~~
r_2=\frac{4}{3\mu}\cos\left(\frac{\theta+2\pi}{3}\right)~~~{\rm and}~~~
r_3=\frac{4}{3\mu}\cos\left(\frac{\theta+4\pi}{3}\right),
\label{roots}
\end{equation}
where
\begin{equation}
\theta=\arccos\left(-\frac{27}{32}\left(M_3+M_4\right)\mu\right). \label{theta-def}
\end{equation}
Since 
$$
-1<-\frac{27}{32}\left(M_3+M_4\right)\mu<0
$$
is satisfied by virtue of Eq.(\ref{UB-muM3M4}), 
\begin{equation}
\frac{\pi}{2}<\theta<\pi \label{theta-restriction}
\end{equation}
holds. Equation (\ref{theta-restriction}) leads to  $r_1>r_3>0>r_2$. 

We introduce a function defined as
$$
U(\rho)=-\frac{4}{3\mu^2}\left[\rho^2-\frac{9(M_3+M_4)}{8}\rho+Q^2\right].
$$
Then, since $r_i$ satisfies 
$$
r_i^4=\frac{4}{3\mu^2}r_i^2-\frac{M_3+M_4}{2\mu^2}r_i,
$$
we have
$$
w(r_i)=U(r_i). 
$$
Because of Eqs.~(\ref{Q-restriction}) and (\ref{LB-M3M4}), the quadratic equation $U(\rho)=0$ has 
two real roots; 
$$
\rho=\rho_\pm :=\frac{9}{16}\left[M_3+M_4\pm \sqrt{(M_3+M_4)^2-\left(\frac{16Q}{9}\right)^2}\right].
$$
If the inequalities,
$$
U(r_1)<0,~~~U(r_2)<0~~~{\rm and}~~~U(r_3)>0, 
$$ 
or equivalently, 
\begin{equation}
r_1>\rho_+,~~~~r_2<\rho_-~~~{\rm and}~~~\rho_-<r_3<\rho_+ \label{condition}
\end{equation}
are simultaneously satisfied, the quartic equation $w(r)=0$ has four real roots. 
We will see below that Eq.~(\ref{condition}) holds. 

Since both $\rho_\pm$ are positive, $r_2<\rho_-$ is trivially satisfied because of $r_2<0$. 

\begin{figure}
\begin{center}
\includegraphics[width=0.7\textwidth]{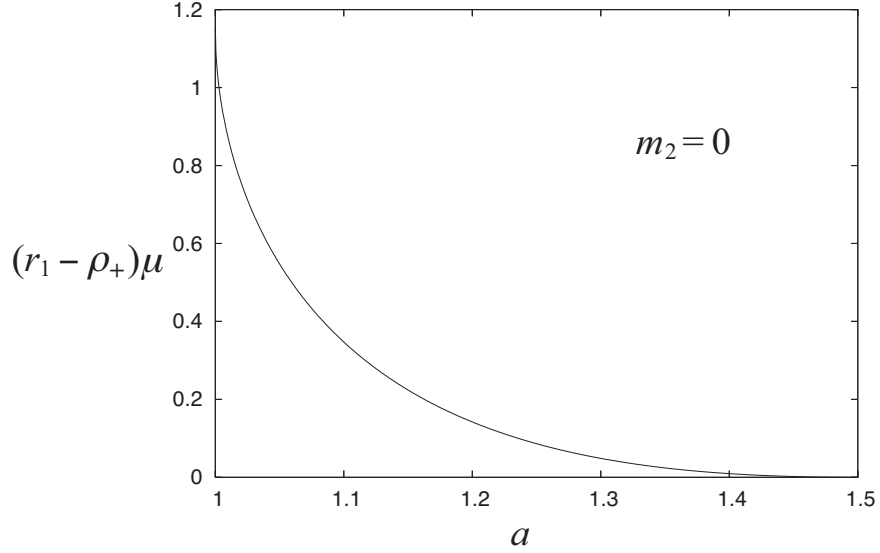}
\caption{\label{fig:radius}
We depict $(r_1-\rho_+)\mu$ with $m_2=0$ as a function of $a$ in the unit of $\Mw=1$.
}
\end{center}
\end{figure}

From Eq.~(\ref{theta-def}), we can see 
$$
\frac{\partial\theta}{\partial m_2}=-\frac{27\mu m_2}{16a\sin\theta} <0,
$$
where we have used Eq.~(\ref{theta-restriction}) in the inequality. Thus, we see 
$$
\frac{\partial r_1}{\partial m_2}=-\frac{4}{9\mu}\sin\left(\frac{\theta}{3}\right)\frac{\partial\theta}{\partial m_2}>0,
$$
and
\begin{equation}
r_1 >r_1|_{m_2=0}=\frac{4}{3\mu}\cos\left[\frac{1}{3}\arccos\left(-\frac{27}{16}\Mw\mu\right)\right]
\label{r1-restriction}
\end{equation}
It is not difficult to see that 
$$
\frac{\partial \rho_+}{\partial m_2}<0	
$$
holds, and hence we have
\begin{equation}
\rho_+ < \rho_+|_{m_2=0}. \label{rho+-restriction} 
\end{equation}
We depict $(r_1-\rho_+)\mu$ for $m_2=0$ in Fig.~\ref{fig:radius}.  
Since, as shown in Fig.~\ref{fig:radius}, $ r_1>\rho_+$ holds for $m_2=0$, 
we have from Eqs.~(\ref{r1-restriction}) and (\ref{rho+-restriction}) 
$$
r_1>\rho_+~~~{\rm for }~~m_2>0.
$$

We can easily see that $\rho_+$ is an increasing function of $M_3+M_4$, whereas $\rho_-$ is a decreasing function 
of $M_3+M_4$, in the domain $\cal D$ of $E\geq1$. 
Then, Eq.~(\ref{LB-M3M4}) implies
\begin{align}
\rho_+&>\rho_+|_{M_3+M_4=\frac{191}{96}\Mw}
=\frac{9}{16}\left[\frac{191}{96}+ \sqrt{\left(\frac{191}{96}\right)^2-\left(\frac{16Q}{9\Mw}\right)^2}\right]\Mw \cr
&>\frac{9}{16}\left[\frac{191}{96}+ \sqrt{\left(\frac{191}{96}\right)^2-\frac{32}{9}}\right]\Mw>\Mw.
\end{align}
We have
\begin{align}
\frac{1}{4}\frac{dw}{dr}\biggl|_{r=\Mw}&=\frac{1}{a^3\mu^2}\left(\Mw^3a^3\mu^2 -\frac{1}{3}\Mw a^3
-\frac{m_2^2 a^2}{2}\right)<-\frac{\Mw}{3a^3\mu^2}f(a), 
\end{align}
where 
$$
f(a)=a^3-6\Mw^2a+6\Mw^3.
$$
It is easy to see that $f(a)>0$ holds for $a>0$. This result implies that $dw/dr|_{r=\Mw}<0$ holds.  
As a result, we have 
$$
r_3<\Mw<\rho_+,
$$
since $r=r_3$ is the lower bound of the domain of $dw/dr<0$ in $r>0$.

\begin{figure}
\begin{center}
\includegraphics[width=0.7\textwidth]{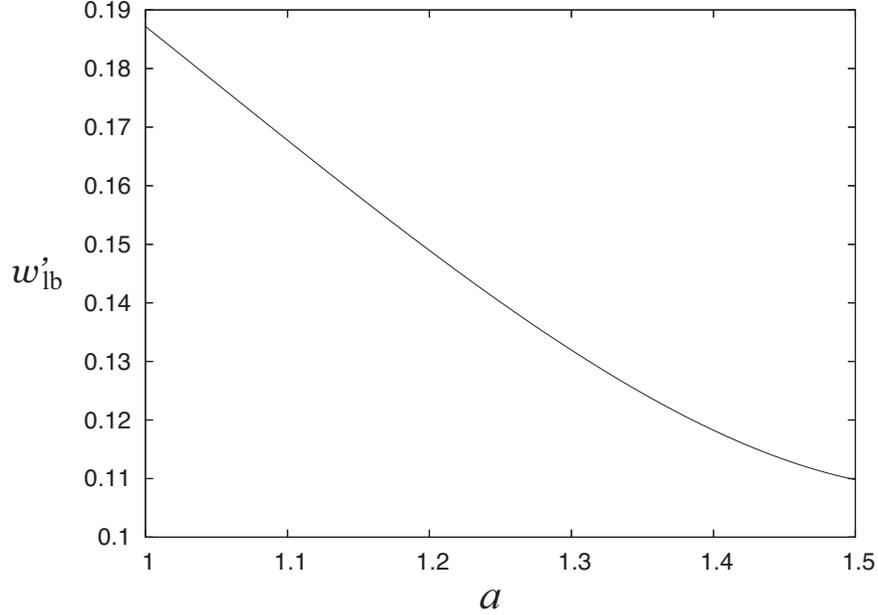}
\caption{\label{fig:radius-2}
We depict $w'_{\rm lb}$ as a function of $a$ in the unit of $\Mw=1$. It is positive in the 
domain of our interest. 
}
\end{center}
\end{figure}

It is easy to see that the following inequality holds in the domain $\cal D$ of $E\geq1$; 
$$
\frac{\partial \rho_-}{\partial m_2}>0,
$$
and hence we have
\begin{align}
\rho_-&<\rho_{\rm ub}(a):=\rho_-|_{m_2=\frac{a-\Mw}{4}}\cr
&=\frac{9}{8}\left[1-\frac{(a-\Mw)^2}{32\Mw a}
- \sqrt{\left(1-\frac{(a-\Mw)^2}{32\Mw a}\right)^2-\left(\frac{8Q}{9\Mw}\right)^2}\right]\Mw
\end{align}
where we have used Eq.~(\ref{no-BH}) and $E\geq1$ in the inequality. 
We can see 
\begin{align}
\frac{\mu^2}{4}\frac{dw}{dr}\biggl|_{r=\rho_{\rm ub}}&
=\mu^2 \rho_{\rm ub}^3-\frac{4}{3}\rho_{\rm ub}+\Mw-\frac{m_2^2}{2a} \cr
&>w'_{\rm lb}(a):=\mu^2 \rho_{\rm ub}^3-\frac{4}{3}\rho_{\rm ub}+\Mw-\frac{(a-\Mw)^2}{32a},
\end{align}
where we have used Eq.~(\ref{no-BH}) and $E\geq1$ in the inequality. 
In Fig.~\ref{fig:radius-2}, we depict $w'_{\rm lb}$ as a function of $a$.  
From this figure, we find that $dw/dr>0$ at $r=\rho_{\rm ub}$, and hence 
$r_3>\rho_{\rm ub}>\rho_-$ holds by the same reason as that leading to $r_3<\Mw<\rho_+$. 
As a result, we have $\rho_-<r_3<\rho_+$.

The result obtained above 
implies that the quartic equation $w(r)=0$ has four real roots. Here recall that  
the function $w(r)$ has one minimum in $r<0$, whereas one maximum 
and one minimum exist in $r>0$. Then, since 
$w<0$ and $dw/dr>0$ at $r=0$, we find that one root of $w(r)=0$ is negative and the other 
three are positive.

\end{document}